%% file: main.tex
\documentclass[sigplan,nonacm]{acmart}

\usepackage{amsmath}
\usepackage[noend]{algpseudocode}
\usepackage[utf8]{inputenc}
\usepackage[ruled,vlined,linesnumbered]{algorithm2e}
\usepackage{enumitem}
\usepackage[textsize=tiny,textwidth=0.6in]{todonotes}
\usepackage{subcaption}

\newcommand{\allnotes}[1]{}
\renewcommand{\allnotes}[1]{#1} 

\usepackage{listings}
\newcommand{\code}[1]{\texttt{\hyphenchar\font=`\_\relax #1}}
\usepackage{pifont}

\newcommand{\MyPara}[1]{\noindent{\bf #1}~~}
\newcommand{\Insight}[1]{{\bf #1}~~}

\newcommand{\sm}{{SM Faults}\xspace}
\newcommand{\mmu}{{MMU Faults}\xspace}
\newcommand{\device}{{Device Faults}\xspace}

\usepackage{multirow}
\usepackage{colortbl}
\usepackage{xcolor}
\usepackage{booktabs}
\usepackage{makecell}
\usepackage{hyperref}
\hypersetup{
  colorlinks=true,
  urlcolor=blue,
  linkcolor=blue,
  citecolor=blue,
}
\definecolor{unreachable}{gray}{0.90}

\usepackage{xcolor}

\newif\ifcleanversion
\cleanversionfalse

\long\def\recovery#1{%
    \ifcleanversion
    \else
        \begingroup
        #1%
        \endgroup
    \fi
}

\newif\ifrevisionversion
\revisionversionfalse

\newif\ifrevisionpassversion
\revisionpassversionfalse
\long\def\revisionpass#1{%
    \ifrevisionpassversion
        \begingroup
        \color{purple}
        #1%
        \endgroup
    \else
        \begingroup
        #1%
        \endgroup
    \fi
}

\begin{document}
\input{_utils}
\title{Characterization-Guided GPU Fault Resilience in NVIDIA MPS}

\renewcommand{\authorsaddresses}{}

\author{Rixin Liu\textsuperscript{1}, Xingqi Cui\textsuperscript{1}, Kaijian Wang\textsuperscript{1}, Xinheng Ding\textsuperscript{2}, Zirui Liu\textsuperscript{2}, Yuke Wang\textsuperscript{1}, Jiarong Xing\textsuperscript{1}}
\affiliation{%
  \institution{\textsuperscript{1}Rice University \quad \textsuperscript{2}University of Minnesota}
  \country{}
}

\input{00_Abstract}

\maketitle

\input{01_Introduction}

\input{02_Backgound_and_Related_Works}

\input{03_Motivation}

\input{04_Tech_Parts}

\input{06_Evaluation}

\input{07_Related_Works}

\input{08_Discussion_and_Conclusion}

\bibliographystyle{ACM-Reference-Format}
\bibliography{refs}

\clearpage
\appendix
\input{appendix}

\end{document}

%% file: _utils.tex
\def \X {paper name}
\newcommand{\lrx}[1]{\textcolor{blue}{#1}}

%% file: 00_Abstract.tex
\begin{abstract}
NVIDIA Multi-Process Service (MPS) enables fine-grained GPU sharing by allowing multiple processes to execute concurrently on the same GPU, making it an important mechanism for improving GPU utilization.
However, MPS has weak fault resilience: a fault in one process can terminate all co-running processes, limiting its adoption in resilience-critical settings such as multi-tenant GPU clusters.
In this work, we design fault-resilient MPS to solve this problem.
Our design is guided by insights from a systematic characterization of GPU faults and a deep analysis of their end-to-end processing pipeline.
Based on these insights, we design two complementary mechanisms.
First, we design a fault isolation mechanism for the dominant memory-related faults that can be fully isolated \revisionpass{while preserving process-level fail-stop semantics} by software intervention in the open GPU driver kernel module.
For other faults whose process is within proprietary software, we design \revisionpass{a fast-recovery substrate that combines virtual-memory-based GPU-resident state sharing with pre-initialized standbys.}
\revisionpass{Our evaluation across GPUs and workloads demonstrates effective fault isolation and fast recovery with minimal overhead: in an end-to-end case study, isolation incurs no visible outage, while recovery restores pre-fault throughput in 355\,ms.}

\end{abstract}

%% file: 01_Introduction.tex
\section{Introduction}

NVIDIA Multi-Process Service (MPS)~\cite{nvidia_mps} enables concurrent GPU execution across processes with configurable Streaming Multiprocessor (SM) usage limits.
Such fine-grained sharing is becoming increasingly important for many scenarios, such as small model serving~\cite{phi,mobilellm} and interactive ML notebook workloads~\cite{notebookos}, where rapidly growing GPU capacity makes full-device allocation inefficient and GPU utilization difficult to manage~\cite{msr_gpu_util, antman, weng2022mlaas}.

However, MPS's weak fault resilience limits its adoption in resilience-critical settings, such as multi-tenant GPU clusters.
MPS collapses the usual CUDA context boundary by multiplexing all processes (MPS clients) into a single shared context, which contains GPU execution state, such as address spaces, kernels, memory allocations, and streams.
As a result, a fatal fault from one client can destroy the shared context and terminate all co-running clients, regardless of which client caused the fault.
In multi-GPU workloads, such as tensor-parallel inference~\cite{megatron}, the failure can further propagate across devices, bringing down execution on all participating GPUs.

This all-or-nothing failure mode forces operators to choose between utilization and resilience.
To obtain stronger resilience, deployments either replace MPS with isolation-oriented mechanisms such as exclusive GPU allocation or fixed-size partitioning via Multi-Instance GPU (MIG)~\cite{nvidia_mig}, or compensate for MPS failures with service-level redundancy such as redundant computation and backup replicas~\cite{bamboo, oobleck}.
Both approaches sacrifice the benefits of fine-grained sharing: MIG can strand capacity due to rigid partition sizes, while standby replicas consume substantial GPU memory due to duplicated model weights and runtime state.
This motivates a system that preserves MPS-like fine-grained sharing while providing strong fault resilience.

Realizing this goal requires answering several fundamental questions:
What GPU faults can occur under MPS?
How are they processed across the GPU hardware and software stack?
Can the stack be enhanced to confine faults to the faulting client through software intervention?
And when isolation is infeasible, can recovery be made fast enough to minimize disruption?
To the best of our knowledge, no prior work has explored these questions.

Until recently, answering these questions was impossible because NVIDIA GPUs expose only narrow, high-level abstractions to applications, while fault handling spans hardware, firmware, and proprietary driver components.
The open-sourcing of NVIDIA's GPU kernel modules~\cite{open_gpu_kernel_modules} creates a new opportunity to investigate the fault-handling implementations.
However, this still remains a challenging job because the relevant fault-handling logic remains largely undocumented and spans deep, recursive call chains across multiple software layers, requiring substantial efforts to analyze systematically and comprehensively.

\begin{table}[t]
\centering
\footnotesize
\caption{GPU fault categories and resiliency solutions.}
\label{tab:overview}
\begin{tabular}{lcllc}
\toprule
\textbf{Category} & \textbf{Count} & \textbf{Handler} & \textbf{Solution} \\
\midrule
\mmu    & 14 & Open-source UVM      & Isolation (\S\ref{sec:isolation}) \\
\sm     & 5  & Closed-source RM/GSP & Fast Recovery (\S\ref{sec:recovery}) \\
\device & -- & Full GPU reset        & Out of scope \\
\bottomrule
\end{tabular}
\end{table}

In this work, we first present a systematic GPU fault characterization (\S\ref{sec:characterization}).
It classifies 19 GPU fault scenarios using two principles based on their fault raisers and properties.
At the top level, these faults fall into three categories---Memory Management Unit (MMU) Faults, \sm, and \device---as summarized in Table~\ref{tab:overview}.
We further trace their end-to-end processing path, from detection to fatality determination and propagation.
This analysis yields four design insights for fault-resilient MPS.
The key insight is that MMU faults can be fully isolated by enhancing the open-source kernel modules, whereas SM faults are processed largely by closed-source firmware, making fast recovery the practical resilience strategy.

Guided by this characterization, we first design a fault isolation mechanism for MMU faults. 
\revisionpass{
The challenge here is to intercept a fatal fault and terminate only the faulting client without leaving the GPU stack inconsistent.
}
Our insight is that NVIDIA's Unified Virtual Memory (UVM) module~\cite{open_gpu_kernel_modules}, originally designed to support GPU page faults for memory oversubscription, 
provides the visibility and control needed for such interception.
Leveraging UVM, we redirect fatal accesses to dummy pages before propagation, making the fault appear resolved from the hardware's perspective and leaving no inconsistent GPU state.
\revisionpass{
We couple this fault resolution with preemption to suspend all GPU execution of the faulting process before it can resume, enabling safe termination while preserving process-level fail-stop semantics.
}
Together with client-granularity termination, this confines the fault to the faulting client while allowing co-running clients to continue execution.
This mechanism isolates all MMU faults, which are the most frequent class of GPU faults, accounting for 94\% in a recent resilience study of over 2.1 million H100 GPU hours~\cite{story-two-gpus}.

We next design a fast recovery mechanism for SM faults.
The key goal is to restore execution as quickly as possible to minimize service disruption.
\revisionpass{
Therefore, this excludes cold restart approaches that reload GPU-resident states and recompute completed work to regain pre-fault progress.
}
\revisionpass{
We avoid such overhead by retaining GPU-resident state across process failures.
Specifically, we use NVIDIA's Virtual Memory Management (VMM) API~\cite{cuda_vmm} to map the same physical GPU memory into both the active and a light-weight standby (backup) processes, allowing the standby to reuse the preserved state without reloading or recomputing it.
}
\revisionpass{
We pair this state reuse with pre-initialized standbys outside MPS, avoiding repeated runtime setup.
Application recovery protocols specify which state can be reused and how execution resumes.
Together, these mechanisms enable millisecond-level recovery across  applications.
}

We implement fault isolation by modifying the open-source UVM kernel module, 
\revisionpass{
and implement the recovery substrate using a build-time \texttt{libcuda.so.1} interceptor and user-space orchestration.
}
To trigger GPU faults efficiently and deterministically for testing, we also develop a fault-injection module (\S\ref{app:fault-injection}).
\revisionpass{
Our evaluation on NVIDIA A6000 and H100 GPUs shows that isolation contains all propagating user-reachable MMU fault scenarios.
In end-to-end case studies, isolation incurs no visible outage, while recovery restores pre-fault throughput in 355\,ms.
Experiments with vLLM~\cite{vllm}, diffusion~\cite{diffusion}, and ResNet50~\cite{resnet} also demonstrate the generality of our fast recovery across application protocols.
}

%% file: 02_Backgound_and_Related_Works.tex
\section{Background}

\begin{figure}[t]
    \centering
    \includegraphics[width=1\linewidth]{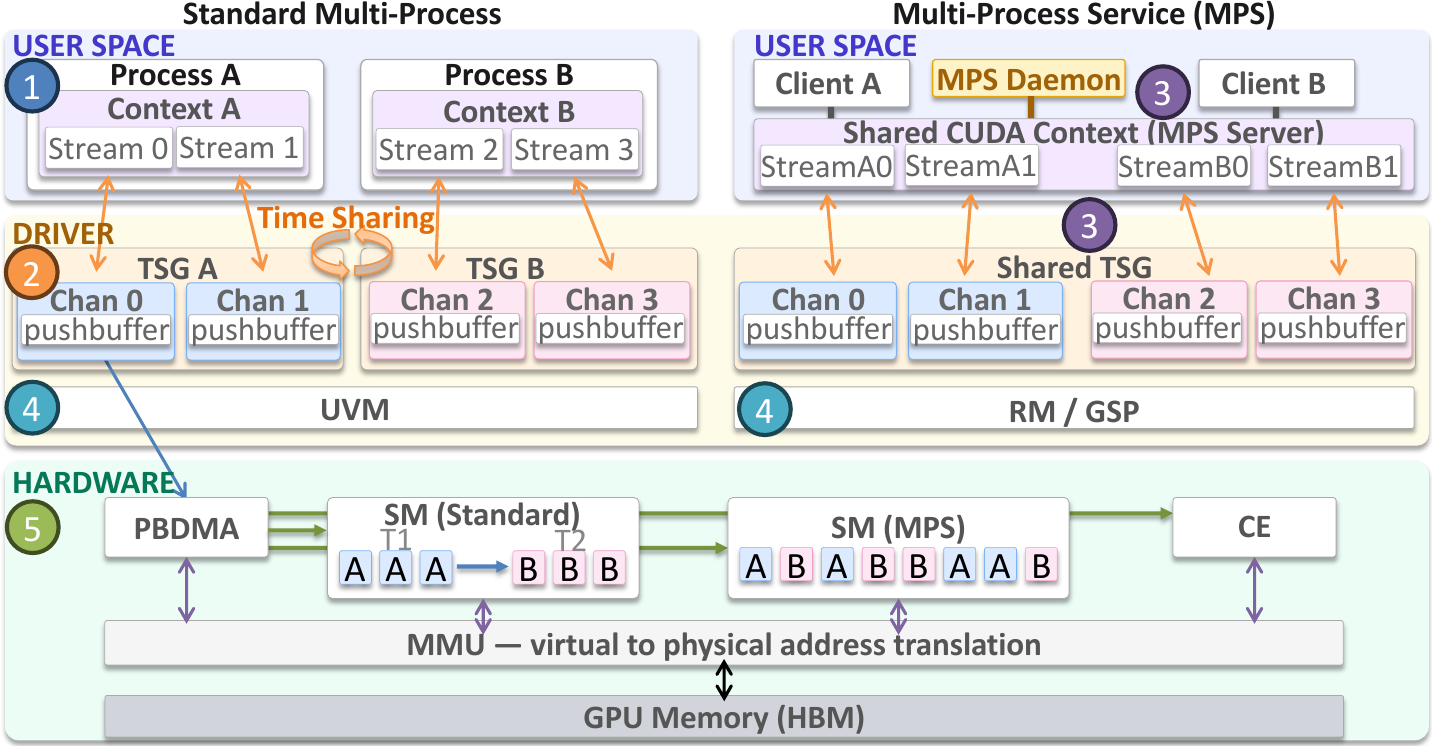}
    \caption{Illustration of the GPU execution model.}
    \label{fig:execution_model}
\end{figure}


\subsection{GPU Execution Model}

Figure~\ref{fig:execution_model} illustrates NVIDIA GPUs' execution model.

\MyPara{User space (\ding{172}).}
A GPU application interacts with the hardware through the CUDA runtime, which manages GPU execution. 
The key abstraction is the \textit{CUDA context}: it encapsulates the process's GPU state and serves as the unit of user-level GPU resource management.
Within a context, applications issue work through \textit{streams}, First-In, First-Out (FIFO) command queues whose operations execute in submission order; operations in different streams may run concurrently.

\MyPara{Channels and TSGs (\ding{173}).}
Each stream maps to a \textit{channel}, whose pushbuffer stages commands to the GPU. 
Channels are grouped into a \textit{Time Slice Group} (TSG), and the GPU hardware scheduler round-robins time slices across TSGs. 
During a TSG's time slice, the scheduler scans its channels for pending commands and may dispatch work from multiple channels concurrently for spatial parallelism.


\MyPara{\textbf{MPS (\ding{174}).}}
Under conventional multi-process execution, each process owns an isolated CUDA context with its own TSG (Figure~\ref{fig:execution_model}, left), so different processes' kernels can only alternate through time-slice rotation and never truly execute concurrently.
MPS~\cite{nvidia_mps} overlays this model through a server-client architecture managed by a background \textit{MPS daemon}, which launches a persistent \textit{MPS server} and exposes a control interface for client management.
The MPS server multiplexes all clients' channels into a shared CUDA context and a single TSG (Figure~\ref{fig:execution_model}, right).
This places channels from different processes in the same TSG, allowing their kernels to be dispatched concurrently within one time slice.
In effect, MPS turns cross-process time-division multiplexing into intra-context spatial sharing.


\MyPara{\textbf{Driver modules (\ding{175}).}}
The scheduling structures above, including channels and TSGs, are created and managed by the GPU driver.
Two components are particularly relevant.
The \textit{UVM module} (\texttt{nvidia-uvm}) is an open-source kernel module responsible for GPU memory management, e.g., page-fault handling and page migration.
The \textit{Resource Manager} (RM) implements hardware resource management, e.g., the creation, destruction, and scheduling configuration of channels and TSGs; much of this logic runs as closed-source proprietary firmware on the on-chip \textit{GPU System Processor} (GSP).


\MyPara{Hardware engines (\ding{176}).}
Once the driver configures these structures, the hardware executes work as follows: 
the \textit{Push Buffer DMA} (PBDMA) unit serves as the host interface, reading commands from the selected channel's pushbuffer, parsing them, and dispatching them to the appropriate GPU engine for execution:
the Compute/Graphics Engine, composed of an array of \textit{Streaming Multiprocessors} (SMs), executes computation kernels, while the \textit{Copy Engine} (CE) handles data movement between GPU memory, CPU memory, and peer devices.
All engines access GPU memory through the \textit{Memory Management Unit} (MMU), which translates virtual addresses to physical addresses.


\subsection{GPU Memory Model}
\label{subsec:gpu_mem}

\recovery{\MyPara{\texttt{cudaMalloc} and VMM APIs.}
\texttt{cudaMalloc} is the traditional and most common GPU memory allocation interface for device-resident memory.
It combines physical allocation and virtual mapping in one operation, leaving applications with no control over the underlying mapping.
The VMM API~\cite{cuda_vmm} separates these steps: \texttt{cuMemCreate} creates a physical allocation, while \texttt{cuMemMap} maps it into a virtual address range.
This decoupling enables zero-copy cross-process sharing, since the same physical allocation can be mapped into multiple processes' address spaces and remains valid while any mapping or handle still references it.}

\MyPara{Unified Virtual Memory (UVM).}
UVM managed memory, allocated through \texttt{cudaMallocManaged}, provides a unified address space accessible by both the CPU and GPU.
Unlike \texttt{cudaMalloc}, it initially reserves only a virtual address range; physical pages are allocated and mapped lazily on first access.
When the GPU touches an unmapped address, the MMU raises a page fault, and the UVM driver allocates and maps a physical page before resuming execution.
If the page is resident in CPU memory, UVM first migrates it to GPU memory via DMA.
Regardless of allocation method, UVM tracks all memory regions as \textit{VA ranges} and classifies them by the level of fault handling they receive.
Memory allocated through \texttt{cudaMalloc} or the VMM API is registered as  \textit{external ranges}, for which UVM does not service page faults.
Only \texttt{cudaMallocManaged} memory is registered as \textit{managed ranges}, which receive full page fault handling support.


%% file: 03_Motivation.tex
\section{Motivation}
\label{sec:motivation}

\subsection{The Need for Fine-Grained GPU Sharing}

Low GPU utilization remains a persistent challenge in modern AI infrastructure, driven by two structural trends.
First, GPU capacity is scaling rapidly, from 80~GB of HBM and multi-PFLOP compute in H100 GPUs~\cite{nvidia-h100-whitepaper} to hundreds of GB of HBM and tens of PFLOPS in next-generation GPUs~\cite{nvidia-blackwell-brief}.
Second, AI workloads are becoming more diverse and heterogeneous, small language models are widely deployed~\cite{phi,mobilellm}, inference traffic is bursty and unpredictable~\cite{qoserve}, and long-tail RL workloads often exhibit irregular execution and low effective parallelism~\cite{gmi, reason}.  
Together, these trends make it difficult for many workloads to fully utilize modern GPUs. 
As a result, this demand has motivated many GPU-sharing systems built on NVIDIA's native MPS~\cite{nvidia_mps} and MIG~\cite{nvidia_mig}, as described in \S\ref{sec:related}.

\subsection{The Sharing-Isolation Dilemma}
\label{subsec:dilemma}

However, today's GPU sharing faces a fundamental tradeoff between utilization and fault isolation. 

\MyPara{High Utilization, No Isolation:} 
MPS~\cite{nvidia_mps} enables fine-grained SM sharing by allowing multiple client processes to share GPU execution resources, improving utilization through tighter workload packing.
However, MPS multiplexes clients within a shared CUDA context.
When a fatal fault occurs, this shared context can be destroyed, terminating all co-running clients regardless of which client caused the fault.
Achieving resilient GPU sharing under MPS therefore requires external redundancy, such as backup services, which can double GPU resource usage and undermine the goal of GPU sharing.

\MyPara{Full Isolation, Low Utilization.}
MIG~\cite{nvidia_mig} partitions a GPU into hardware-isolated instances, providing strong fault isolation.
However, MIG offers limited partition flexibility: a GPU supports at most seven instances, with only a small set of fixed instance sizes.
Workloads whose resource demands do not match these predefined sizes can leave substantial GPU capacity stranded.
Moreover, MIG partitions cannot be reconfigured without draining running workloads, making it poorly suited to environments with fluctuating demand.


To break this dilemma, this paper asks whether MPS can support fine-grained GPU sharing with fault resilience\footnote{Starting with driver r610, MPS supports fault isolation across static SM partitions~\cite{new-mps}.
However, this technology itself represents a trade off between utilization and isolation. See more discussion in \S\ref{sec:discussion}.}: fault isolation where possible with fast recovery where necessary.
To the best of our knowledge, no prior work has explored this question.
The closest attempt is MuxFlow~\cite{muxflow}, which identifies faults caused by incorrect client termination under MPS.
However, it does not address faults that arise during GPU execution, such as illegal memory accesses.

\subsection{Our Work: Fault-Resilient MPS}

We break this sharing--isolation dilemma by making MPS fault-resilient, preserving fine-grained GPU sharing while adding strong fault resiliency.
This broadens MPS-based GPU sharing to resiliency-critical deployments, enabling operators to improve utilization without sacrificing fault resiliency.

This work includes three components.
First, we systematically characterize GPU faults under MPS, classifying 19 fault scenarios and tracing their end-to-end processing paths from detection to fatality determination and propagation (\S\ref{sec:characterization}).
This provides insights for our fault isolation and fast recovery design.
Second, we design a UVM-based fault isolation mechanism for MMU faults, intercepting them in the open GPU kernel module and confining their impact to the faulting client (\S\ref{sec:isolation}).
\recovery{Third, for SM faults that are handled inside proprietary components and therefore difficult to intercept directly, we design a VMM-based active-standby recovery mechanism that reuses GPU-resident state to restore execution in milliseconds (\S\ref{sec:recovery}).}

%% file: 04_Tech_Parts.tex
\section{GPU Fault Characterization}
\label{sec:characterization}


\begin{table*}[t]
\centering
\scriptsize
\setlength{\tabcolsep}{4pt}
\renewcommand{\arraystretch}{1.15}
\caption{\revisionpass{GPU fault taxonomy validated on NVIDIA \textit{RTX A6000, L40, A100, and H100 GPUs}.
Gray cells denote scenarios unreachable through CUDA APIs; \texttt{ioctl} denotes injection through UVM debug interfaces.}}
\label{tab:memory-fault-taxonomy}
\begin{tabular}{llclllcccc}
\toprule
\textbf{Category} & \textbf{\#} & \textbf{Replayability.} & \textbf{Fatality Stage} & \textbf{Serviceability} & \textbf{Fault Scenario} & \makecell[c]{\textbf{Faulting}\\\textbf{Engine}} & \makecell[c]{\textbf{Reachability}} & \makecell[c]{\textbf{Propagates}} & \textbf{Solution} \\
\midrule
\multirow{22}{*}{\textbf{MMU Faults}}
 & \cellcolor{unreachable} & \multirow{10}{*}{{Replayable}}
       & \cellcolor{unreachable}\makecell[l]{Parse-time} & \cellcolor{unreachable}\makecell[l]{Non-serviceable} & \cellcolor{unreachable}HW error conditions & \cellcolor{unreachable}SM & \cellcolor{unreachable}unreachable & \cellcolor{unreachable}n/a & \cellcolor{unreachable}N/A \\
\cmidrule(l){2-2}\cmidrule(l){4-10}
 & 1   & & \multirow{8}{*}{\makecell[l]{Deferred to servicing}} & \multirow{6}{*}{\makecell[l]{Non-serviceable}} & Out-of-bounds access (OOB) & SM & \textbf{Yes} & \textbf{Yes} & M1 (\S\ref{sec:isolation}) \\
 & 2   & & & & Access mismatch (CPU-resident) & SM & \textbf{Yes} & \textbf{Yes} & M2 (\S\ref{sec:isolation}) \\
 & 3   & & & & Access mismatch (GPU-resident) & SM & \textbf{Yes} & \textbf{Yes} & M2 (\S\ref{sec:isolation}) \\
 & 4   & & & & Access mismatch (VMM memory) & SM & \textbf{Yes} & \textbf{Yes} & M3 (\S\ref{sec:isolation}) \\
 & 5   & & & & Zombie range access & SM & ioctl & \textbf{Yes} & M2 (\S\ref{sec:isolation}) \\
 & 6   & & & & Non-migratable range access & SM & ioctl & \textbf{Yes} & M2 (\S\ref{sec:isolation}) \\
\cmidrule(l){5-10}
 &     & & & \multirow{2}{*}{\makecell[l]{Serviceable\\(benign)}} & Page fault (demand paging) & SM & normal & No & N/A \\
 &     & & & & Invalid prefetch & SM & normal & No & N/A \\
\cmidrule(l){2-10}
 & \cellcolor{unreachable} & \multirow{12}{*}{{Non-replayable}}
       & \cellcolor{unreachable}\makecell[l]{Parse-time} & \cellcolor{unreachable}\makecell[l]{Non-serviceable} & \cellcolor{unreachable}HW error conditions & \cellcolor{unreachable}CE/PBDMA & \cellcolor{unreachable}unreachable & \cellcolor{unreachable}n/a & \cellcolor{unreachable}N/A \\
\cmidrule(l){2-2}\cmidrule(l){4-10}
 & 7   & & \multirow{8}{*}{\makecell[l]{Deferred to servicing}} & \multirow{6}{*}{\makecell[l]{Non-serviceable}} & OOB & CE & \textbf{Yes} & Contained & N/A \\
 & 8   & & & & Access mismatch & CE & \textbf{Yes} & Contained & N/A \\
 & \cellcolor{unreachable}9   & & & & \cellcolor{unreachable}Zombie range & \cellcolor{unreachable}CE & \cellcolor{unreachable}unreachable & \cellcolor{unreachable}n/a & \cellcolor{unreachable} \\
 & \cellcolor{unreachable}10  & & & & \cellcolor{unreachable}Non-migratable & \cellcolor{unreachable}CE & \cellcolor{unreachable}unreachable & \cellcolor{unreachable}n/a & \cellcolor{unreachable} \\
 & 11  & & & & OOB & PBDMA & \textbf{Yes} & \textbf{Yes} & M1 (\S\ref{sec:isolation}) \\
 & \cellcolor{unreachable}12  & & & & \cellcolor{unreachable}\textit{Access mismatch} & \cellcolor{unreachable}PBDMA & \cellcolor{unreachable}unreachable & \cellcolor{unreachable}n/a & \cellcolor{unreachable} \\
 & \cellcolor{unreachable}13  & & & & \cellcolor{unreachable}\textit{Zombie range} & \cellcolor{unreachable}PBDMA & \cellcolor{unreachable}unreachable & \cellcolor{unreachable}n/a & \cellcolor{unreachable} \\
 & \cellcolor{unreachable}14  & & & & \cellcolor{unreachable}\textit{Non-migratable} & \cellcolor{unreachable}PBDMA & \cellcolor{unreachable}unreachable & \cellcolor{unreachable}n/a & \cellcolor{unreachable} \\
\cmidrule(l){5-10}
 &     & & & \multirow{2}{*}{\makecell[l]{Serviceable\\(benign)}} & Page fault & CE & normal & No & N/A \\
 &     & & & & Page fault & PBDMA & normal & No & N/A \\
\midrule
\multirow{5}{*}{\makecell[l]{\textbf{SM Faults}}}
 & 15 & \multicolumn{3}{l}{\multirow{5}{*}{\makecell[l]{Handled entirely within closed-source\\RM/GSP firmware. Discovery methodology\\ is detailed in \S\ref{app:fault-injection}.}}}
   & Lane user stack overflow (EXC\_2) & SM & \textbf{Yes} & \textbf{Yes} & \S\ref{sec:recovery} \\
 & 16 & & & & Illegal instruction (EXC\_4) & SM & \textbf{Yes} & \textbf{Yes} & \S\ref{sec:recovery} \\
 & 17 & & & & Shared/local OOB (EXC\_5) & SM & \textbf{Yes} & \textbf{Yes} & \S\ref{sec:recovery} \\
 & 18 & & & & Misaligned address (EXC\_6) & SM & \textbf{Yes} & \textbf{Yes} & \S\ref{sec:recovery} \\
 & 19 & & & & Invalid address space (EXC\_7) & SM & \textbf{Yes} & \textbf{Yes} & \S\ref{sec:recovery} \\
\midrule
\textbf{Device Faults} & & \multicolumn{4}{l}{Thermal faults, device failures. GPU entirely unavailable.} & \multicolumn{3}{c}{\textit{Full device reset required. Out of scope.}} \\
\bottomrule
\end{tabular}
\end{table*}

Although NVIDIA open-sources its GPU kernel modules, the fault-handling logic is not externally documented, and the relevant control flow spans deep, recursive call chains across multiple software layers.
To provide a knowledge foundation for designing fault-resilient MPS, this section characterizes GPU faults through a systematic study of NVIDIA's GPU fault-handling infrastructure, guided by three questions:
\textbf{Q1}: What types of faults can occur during GPU execution, and how should they be classified?
\textbf{Q2}: How does the driver handle a fault from detection to determination as fatal?
\textbf{Q3}: How does a fatal fault propagate from the faulting client to co-running clients under MPS?


\subsection{Systematic and Comprehensive Fault Taxonomy}
\label{subsec:taxonomy}

We first address Q1 through a systematic and comprehensive taxonomy that classifies GPU faults by two principles into three categories spanning 19 distinct scenarios. 
Table~\ref{tab:memory-fault-taxonomy} summarizes the complete taxonomy.
\begin{itemize}[leftmargin=1.5em, itemsep=0.4em, topsep=0.4em]
    \item \textbf{Principle 1:} \textit{\textbf{By fault raisers,}} faults are partitioned at the top level into three categories. 
    \item \textbf{Principle 2:} \textit{\textbf{By fault properties,}} faults handled by the open-source driver module are further sub-classified based on their properties.  
\end{itemize}

\subsubsection{GPU Faults by Raisers (Principle 1)}\mbox{}
\label{subsubsec:principle1}

\MyPara{Category 1: MMU Faults} are raised when the MMU fails to translate a virtual address during a memory access initiated by any of the three faulting engines: SM, CE, or PBDMA.
These faults are handled entirely within the open-source UVM kernel module, enabling fine-grained observation and modification of the handling path.

\revisionpass{
\MyPara{Category 2: SM Faults}
are raised when an SM encounters an execution exception.
To systematically enumerate fatal SM faults, we exercise the exception codes documented by \texttt{cuda-gdb}~\cite{cuda_gdb} and trace their handling paths, excluding MMU faults and codes that do not produce fatal SM exceptions on the tested GPUs (Appendix~\ref{app:fault-injection}).
This identifies five fault types: lane user stack overflow, illegal instruction, shared/local out-of-range access, misaligned address, and invalid address space.
These faults are handled entirely within closed-source RM/GSP firmware.
}

\MyPara{Category 3: Device Faults} are raised by device-level failures, such as thermal faults or uncorrectable hardware errors, that render the entire GPU unavailable and require a full device reset.
These faults are out of scope for this work.

\subsubsection{MMU Faults by Properties (Principle 2)}\mbox{}
\label{subsubsec:principle2}

\noindent We further classify MMU faults along three fault properties. 
\MyPara{Replayability} indicates whether the hardware will retry to solve the fault.
\revisionpass{
This classification is historical: in pre-Volta architectures, only MMU faults from SMs support hardware retry, while those from CE or PBDMA do not.
Post-Volta architectures restore retry capability to the latter faults, but UVM retains the original classification, still labeling them as ``non-replayable.''
}

\MyPara{Fatality stage} then captures \textit{when} in the pipeline a fault is determined fatal: \textit{parse-time fatal} faults are identified during initial parsing---these correspond to errors that cannot be resolved through any software intervention; 
while \textit{deferred-to-servicing} faults are exposed only when the driver attempts to resolve them.

\MyPara{Serviceability} next distinguishes the outcome of the servicing stage: serviceable faults are resolved silently by the driver, while non-serviceable faults cannot be resolved and proceed toward fatal propagation and program termination.

\MyPara{Fault Scenarios and Reachability.}
As shown in Table~\ref{tab:memory-fault-taxonomy}, the three fault properties partition deferred-to-servicing MMU faults into four base fault conditions: out-of-bounds access (OOB), access mismatch (AM), zombie range access, and non-migratable range access.
AM further splits into three variants for MMU faults: two by page residency (CPU-resident and GPU-resident) and one for externally managed VMM ranges.
Crossing these conditions with the three faulting engines (SM, CE, PBDMA) produces 14 engine-fault scenarios.
\revisionpass{Nine of these scenarios are reachable from user space: seven through CUDA APIs and two (\#5 and \#6) through UVM debug ioctls,} while the remaining five (\#9, \#10, \#12, \#13, \#14) are architecturally unreachable.
Two additional scenarios, demand paging and invalid prefetch, are serviceable and resolved transparently by the driver.
Detailed descriptions of fault scenarios and unreachable combinations are in Appendix~\ref{sec:appendix-faults}.

\subsection{Fault Detection and Fatality Determination}
\label{subsec:pipeline}

We next examine how the GPU detects a fault and determines its fatality, the first two steps in Figure~\ref{fig:fault-lifecycle}.

\MyPara{\ding{182} Detection.}
GPU hardware reports fault signals to the driver through two paths.
(1) SM faults: when an SM encounters an execution exception, the hardware notifies RM/GSP through a global TRAP signal that reports the error type observed on that SM, but carries no channel information.
(2) MMU faults: when virtual address translation fails, the MMU writes a fault packet into a hardware fault buffer, encoding the faulting virtual address, access type, and fault type.

The two paths differ in attribution and buffering.
By studying Nouveau~\cite{nouveau}, the Linux community's reverse-engineered NVIDIA GPU driver, we find that MMU fault packets include the channel ID of the faulting context, whereas the SM global TRAP path provides no per-channel attribution.

Furthermore, within the MMU path, replayable and non-replayable faults use different buffer ownership: UVM directly manages replayable-fault buffers through hardware GET/PUT registers, while RM/GSP owns non-replayable-fault buffers and copies their contents into a shadow buffer before notifying UVM.
Despite this difference, both cases are ultimately handled by UVM, which uses the channel ID to identify the faulting client and perform targeted resolution.

\Insight{\emph{Insight \#1: The per-channel attribution from the MMU path offers the information needed for faulting client identification.}} 


\MyPara{\ding{183} Fatality Determination.}
Once a fault reaches the driver, handling diverges between the two categories.
SM faults are handled entirely inside the closed-source RM/GSP, making their path opaque to external tracing.
We therefore focus our analysis on MMU faults.

Upon receiving an interrupt, UVM retrieves fault packets from the buffer through a two-level model: a top-half Interrupt Service Routine (ISR) reads pending entries and queues a bottom-half work item on a kernel worker thread.
In the bottom half, UVM handles faults according to the three fault properties.
It first classifies the fault by \textbf{replayability}.
For replayable faults, the MMU follows a \textit{fault-and-stall} model: it stalls the faulting GPU channel, along with other channels in the same TSG, and waits for UVM to resolve the fault before issuing a replay command.
For non-replayable faults, the hardware follows a \textit{fault-and-switch} model: it immediately preempts the TSG containing the faulting channel, allowing other TSGs to continue execution.

\revisionpass{
\Insight{\emph{Insight \#2: Both paths interrupt normal execution of the affected TSG, providing an intervention window before fault servicing allows execution to resume.}}
}

UVM then parses each fault entry and checks for \textbf{fatality}.
Some faults are marked fatal during initial parsing when their internal fault type exceeds UVM's fatality threshold (Table~\ref{tab:memory-fault-taxonomy}, ``Parse-time'' rows).
Faults that pass this check proceed to servicing, where UVM determines \textbf{serviceability}.
Serviceable faults are benign memory-management events, such as demand paging and invalid prefetch requests (Table~\ref{tab:memory-fault-taxonomy}, ``Benign'' rows), and are resolved transparently before execution resumes.
Non-serviceable faults cannot be resolved and are marked fatal at this stage (Table~\ref{tab:memory-fault-taxonomy}, rows~\#1--\#8 and~\#11).
\Insight{\emph{Insight \#3: RM/GSP observes a fatal MMU fault only after UVM reports it; by resolving the fault before that report, we can prevent shared-context teardown.}}

\subsection{Fatal Fault Propagation}
\label{subsec:propagation}
\begin{figure}[t]
    \centering
    \includegraphics[width=\linewidth]{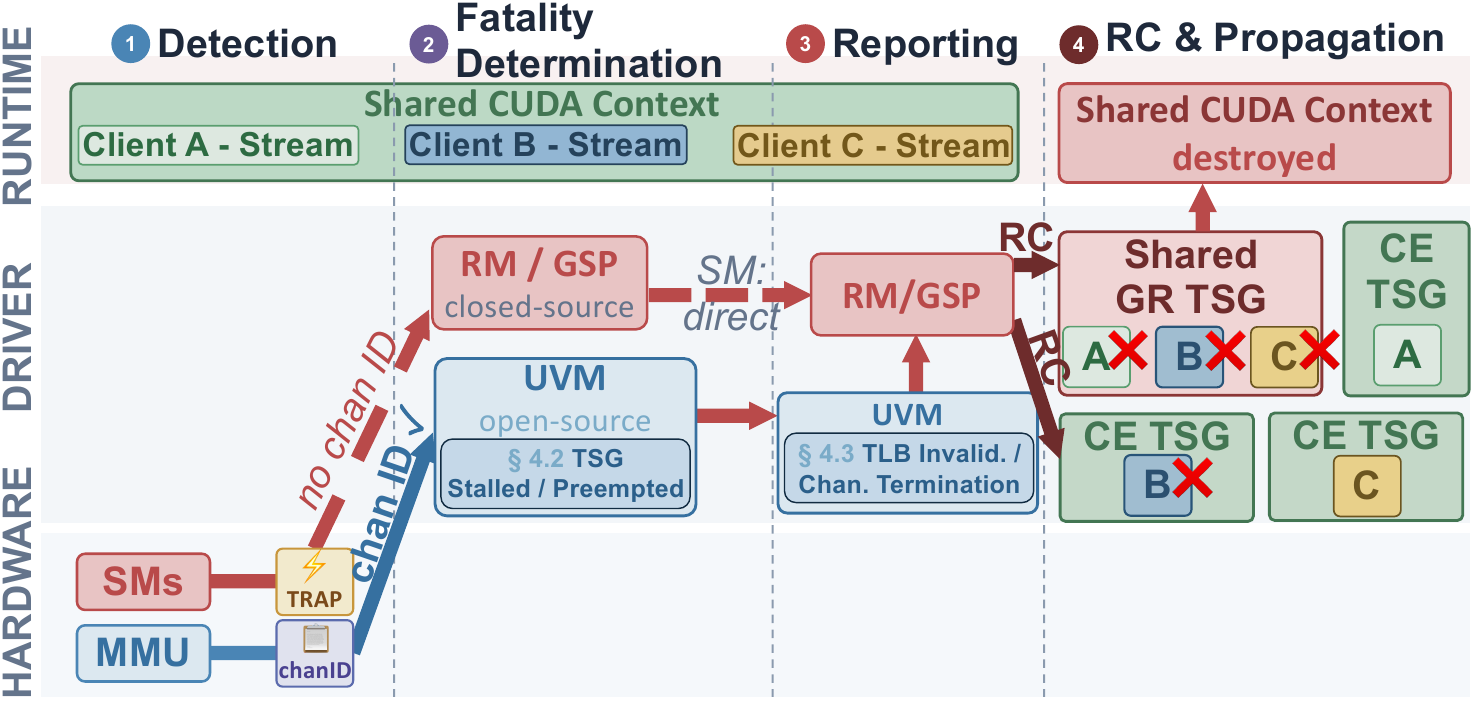}
    \caption{The GPU fault lifecycle and its handling process.}
    \label{fig:fault-lifecycle}
\end{figure}

A key question for fault isolation is whether every reachable fatal fault propagates to all co-running MPS clients.
Among the nine reachable fatal MMU fault combinations (\#1--\#8 and \#11), seven propagate to all co-running clients; all five SM faults also propagate through RC recovery.
Only two MMU combinations, \#7 and \#8 (CE OOB and CE AM), are naturally contained without affecting other clients.

\MyPara{\ding{184} Within UVM: Fatal Fault Reporting.}
Once a fault is classified as fatal (Figure~\ref{fig:fault-lifecycle}, \ding{183}), the driver reports it to the RM/GSP for further action.
For SM faults, handling already occurs inside RM/GSP, so no additional reporting step is needed.
For MMU faults, UVM follows one of two observable reporting paths.
(1) Fatal \textbf{replayable faults}:  UVM issues a Translation Lookaside Buffer (TLB) invalidation command to the faulting GPU channel, after which RM/GSP takes over.
(2) Fatal \textbf{non-replayable faults}: UVM schedules the faulting channel for termination and hands the fault packet to the RM/GSP directly.

\MyPara{\ding{185} Within Driver: RC Recovery and Propagation.}
Upon receiving the fatal report, the RM/GSP performs Robust Channel (RC) recovery, tearing down \textit{all} channels within the affected TSG while other TSGs remain intact.
This coarse granularity stems from the lack of per-channel attribution in the SM fault path (as described in \S\ref{subsec:pipeline}, \ding{182}).

Under MPS, the impact of TSG-level recovery depends on engine-specific sharing policies: 
SM and PBDMA channels from all clients are multiplexed onto a shared GR TSG, whereas each client has an independent CE TSG.
Thus, a fatal fault from an SM or PBDMA channel destroys the shared GR TSG and terminates all clients bound to the shared CUDA context, regardless of which client caused the fault.
In contrast, a fatal fault from an CE channel destroys only the faulting client's CE TSG, naturally containing the error.
In both cases, RM/GSP writes an error notifier that tools such as \texttt{cuda-memcheck} can poll, and the CUDA runtime propagates the error to the application through calls such as \texttt{cudaDevice\allowbreak Synchronize}.

\Insight{\emph{Insight \#4: SM faults lack isolation due to hardware limitations (i.e., no per-channel identity\footnote{NVIDIA's r610 static partitioning enables SM fault attribution through exclusive SM ownership~\cite{new-mps}}}), and the closed-source RM/GSP implementation further complicates software interception and handling, hindering the approach used for MMU faults.} 
This makes fast recovery the only viable strategy for SM faults with today's GPU stack.

\section{MMU Fault Isolation}
\label{sec:isolation}

Building on Insights~\#1--\#3 from \S\ref{sec:characterization}, we achieve complete isolation for all nine reachable MMU fault scenarios (\#1--\#8 and \#11 in Table~\ref{tab:memory-fault-taxonomy}) by enhancing the open GPU kernel modules. 
Software-originated MMU faults are the dominant class of critical GPU errors, accounting for 94\% of reported faults in a recent study of over 2.1 million H100 GPU hours~\cite{story-two-gpus}.
In the following, we first describe the requirement and challenge of isolating MMU faults, and then present our UVM-based interception design.

\subsection{Isolation Requirements and Challenges}
\label{subsec:requirements}

\revisionpass{
\MyPara{Overall Requirement.}
Isolating a fatal MMU fault under MPS requires containing both its hardware and execution effects within the faulting client, while allowing co-running clients to continue their execution.
This requires three capabilities.
First, the fault must be attributable to its originating client.
Second, the fault must be properly resolved without leaving the shared TSG stalled or the driver in an inconsistent state.
Third, fault resolution must preserve process-level fail-stop semantics: once the fault is resolved, the faulting client must remain quiescent until it is safely terminated by the MPS daemon.
The first capability can be achieved leveraging the per-channel identity preserved by UVM (Insight~\#1), which identifies the faulting client.
The latter two capabilities constitute the key design challenges.
}

\MyPara{Challenge 1: Hardware-Safe Fatal Fault Suppression.}
A naive approach is to intercept the fatal-fault report inside UVM before it reaches RC recovery.
As shown in \S\ref{subsec:propagation}, both replayable and non-replayable fatal MMU faults eventually pass through UVM-to-RM/GSP reporting paths.
Suppressing these reports prevents RM/GSP from observing the fatal fault and tearing down the shared CUDA context.
However, suppression alone does not restore the GPU to a safe executable state.
For replayable faults, the MMU stalls the faulting GPU channel and waits for the fault to be resolved.
Suppressing the fatal report only will cause the stalled channel occupying SM resources forever.
Similarly, for non-replayable faults, the hardware preempts the entire TSG that the faulting channel belongs to.
Suppression prevents RC recovery from destroying the shared context, but the TSG remains preempted, blocking all co-running clients whose channels reside on the same TSG.

Therefore, a proper isolation mechanism must first resolve the fault from the hardware's perspective, so that stalled channels can replay or preempted TSG can resume execution, and then cleanly terminate only the faulting client without tearing down the shared CUDA context.

\MyPara{Challenge 2: Preserving Process-Level Fail-Stop Semantics.}
A na\"ive approach is to invoke the MPS daemon's client-termination mechanism after resolving the fault.
The MPS daemon can safely tear down the target client without disrupting the MPS server or co-running clients, even when the target has outstanding GPU work.
However, the target process may run kernels concurrently on multiple CUDA streams.
Although MPS daemon termination prevents the faulting stream from proceeding, kernels on other streams continue executing normally before termination completes.
Therefore, MPS daemon termination provides only stream-level fail-stop semantics, not the process-level guarantee that all streams remain stopped until termination completes.
A proper isolation mechanism must stop the entire faulting process as a single failure unit; otherwise, partial continuation may lead to unpredictable application behavior.

\subsection{Design}
\label{subsec:isolation-design}


\revisionpass{
We intercept faults at UVM's fatality determination point: after UVM has parsed and classified a fault, but before it reports the fault as fatal to RM/GSP and triggers RC recovery. 
We first attribute the fault to its originating MPS client using the per-channel identity preserved by UVM (Insight~\#1). 
Redirect-on-Fault (RoF; \S\ref{subsubsec:rof}) then resolves the fault through UVM's normal servicing path, preventing recovery from escalating to the shared context and addressing Challenge~1. 
To prevent the target process's all streams from continue execution, Preempt-on-Replay (PoR; \S\ref{subsubsec:por}) binds the mandatory replay to a process-level execution barrier, quiescing all GPU streams belonging to the faulting process and addressing Challenge~2. 
Once preemption completes, we invoke the MPS daemon's existing client-termination mechanism to safely tear down the target and re-enable the surviving clients.
}

\subsubsection{Redirect-on-Fault}\mbox{}
\label{subsubsec:rof}

\noindent Within the interception window, if a valid memory mapping is installed before UVM classifies the fault as fatal, UVM will resolve the fault through its normal service path.
From RM/GSP's perspective, no fatal fault has occurred, so RC recovery is never triggered.

This can be achieved by creating \emph{dummy page mappings} through UVM.
Rather than allocating real memory for the faulting address, we redirect it to a shared, dummy page from a driver-global pool.
Because all faults share the same pre-installed dummy page, the mechanism requires no per-fault memory allocation, keeping handling latency minimal.
\revisionpass{
The dummy page is separated from surviving clients' allocations, so redirected accesses cannot read or overwrite their application memory.
}

While the dummy page itself is shared across all cases, the specific mechanism for installing it varies by the state of the virtual address range (\textit{VA range}, \S\ref{subsec:gpu_mem}) at the fault address, yielding three dispatch paths.

\MyPara{Mechanism 1 (M1): Range Creation.} This handles faults where no \textit{VA range} exists at the faulting address, corresponding to out-of-bounds faults such as SM OOB \#1 and PBDMA OOB \#11 in Table~\ref{tab:memory-fault-taxonomy}.
The mechanism creates a new UVM managed range at the fault address and installs a shared 4\,KiB dummy page from a driver-global pool, so UVM services the fault through its normal demand-paging path.

\MyPara{Mechanism 2 (M2): Chunk Substitution.} 
This handles faults where a UVM managed range exists but the target page is not accessible to the faulting processor, such as access-mismatch faults \#2, \#3 in Table~\ref{tab:memory-fault-taxonomy}. 
It redirects the faulting page by replacing its backing chunk with a dummy chunk.
If the original page is GPU-resident, the original chunk is freed in the same pass; if it is CPU-resident, the chunk slot is empty before redirection, so only the dummy chunk is allocated.
In both cases, the final memory footprint matches what UVM would allocate for a legal first-touch operation: exactly one chunk per faulting page.\footnote{\label{fn:shared-backing}
\revisionpass{Reclamation and footprint accounting apply only to unshared backing; shared backing is retained for other clients, and redirection changes only the faulting client's mappings.}}

\MyPara{Mechanism 3 (M3): Range Conversion.} 
This handles faults on external ranges created through the VMM APIs, such as access-mismatch faults (\#4 in Table~\ref{tab:memory-fault-taxonomy}), for which UVM provides no fault servicing. 
To handle these faults, we first destroy the external range and recreate it as a UVM managed range over the same virtual address span.\textsuperscript{\ref{fn:shared-backing}}
Before the service path runs, a shared 2\,MiB dummy chunk from the driver-global pool is installed in the block's chunk array, causing the normal populate logic to short-circuit.

\revisionpass{
\subsubsection{Preempt-on-Replay.}\mbox{}
\label{subsubsec:por}
RoF resolves the hardware fault, but it also returns the faulting process to execution before the MPS daemon terminates it.
The desired behavior is to allow servicing to complete without letting the faulting process resume normal execution.
Our key insight to solve this problem is that a pending preemption cannot complete until the faulting access is replayed, but prevents the TSG from resuming normal instruction execution after replay.
This ordering arises from two complementary constraints. 
The unresolved access prevents the TSG from reaching the architectural state required for context save, whereas an accepted preemption request prevents the TSG from resuming normal execution.

To enforce this ordering, PoR first disables all channels in the shared TSG scheduling, stopping the GPU from fetching new commands in their pushbuffers. 
It then submits an asynchronous TSG preemption request and confirms that the request is pending before modifying the faulting mapping or issuing replay. 
PoR then redirects the faulting access to the dummy page and replays it, allowing the blocked instruction to complete and the pending preemption to proceed. 
If other resident warps expose additional faults before preemption completes, PoR services and replays them while keeping the TSG frozen and the preemption request pending. 
This process continues until the hardware reports preemption completion.

PoR treats preemption completion, rather than replay completion, as the process-wide execution barrier. 
At this point, the shared TSG has been removed from GPU execution, quiescing all streams belonging to the faulting process. 
PoR never re-enables the target channels: it safely tears down the target through the MPS daemon and subsequently resumes only the surviving clients.
}

\section{Fast Recovery for SM Faults}
\label{sec:recovery}

The previous section provided isolation for the dominant MMU fault category.
Following Insight~\#4, we complement that mechanism with fast recovery for SM faults.
We first define the design scope and identify the recovery challenges (\S\ref{subsec:recovery-model}), then present our design (\S\ref{subsec:recovery-design}).

\subsection{Design Scope and Challenges}
\label{subsec:recovery-model}

\MyPara{Scope: Process Failure with a Surviving GPU.}
Prior recovery systems focus on GPU failures that make the device and its resident state unavailable~\cite{dejavu,spotserve,serverlessllm,gemini,megascale}.
We work on a complementary scenario---fatal SM faults under MPS, which invalidate the shared CUDA context but leave the GPU operational.
This creates an opportunity to retain GPU-resident state across process failures and reuse it during recovery.
Device faults requiring a full GPU reset (last row in Table~\ref{tab:memory-fault-taxonomy}) are therefore out of our scope.

\MyPara{Challenge: Avoiding Reinitialization and State Reconstruction.}
Na\"ively restarting a failed process discards reusable GPU-resident state, forcing costly data reloading and recomputation.
This can incur tens of seconds of latency as measured in Table~\ref{tab:recovery-protocols}.
The key challenge is therefore to preserve and reuse this state as much as possible.
Doing so requires overcoming three obstacles.
First, conventional CUDA allocations are tied to their creating process, preventing reuse of even valid data after that process exits.
Second, preserved intermediate state must be accompanied by recovery metadata that identifies where execution can safely resume; otherwise, the application must repeat completed computation.
Third, reusing application state still requires a prepared execution environment: launching a replacement process repeats the initialization process unless that original environment also survives the fault.

\revisionpass{
\subsection{Design}
\label{subsec:recovery-design}

\MyPara{Generic GPU State Recovery Substrate.}
We design an application-independent substrate for preserving and reusing GPU-resident state.
This substrate is compatible and integrable with higher-level recovery orchestration systems such as Kubernetes~\cite{kubernetes}, enabling state reuse when recovering failed processes on a surviving GPU.
These higher-level frameworks can define the application-specific semantics: which state is reusable, what metadata must be preserved, and the execution boundary from which to resume.

Our recovery substrate has three components: VMM-based memory sharing to preserve GPU state, an MPS-aware active--standby architecture for fast failover, and post-recovery lifecycle management to make recovery repeatable.

\MyPara{VMM-Based Memory Reuse.}
CUDA VMM provides the opportunity to manage physical allocations independently of process-local virtual mappings.
We exploit this capability to preserve reusable GPU-resident state across process failures.
Specifically, we intercept CUDA allocation calls and replace them with VMM-backed allocations without changing the application's allocation interface.
For application-designated state, we export shareable allocation handles and map the same physical pages into both the active and standby processes.
The standby holds an independent reference to each allocation, keeping these pages allocated after the active process exits.
Recovery can then reuse the pages directly in the standby or map them into a replacement process, avoiding data copying and reloading.
Together with the required recovery metadata, retained intermediate state also enables the application to resume from its recovery boundary without repeating the corresponding computation.

\MyPara{MPS-Aware Active--Standby Architecture.}
We place each active process under MPS for spatial sharing and its standby in an independent CUDA context outside MPS.
We initialize the standby in advance, then keep it quiescent without launching GPU kernels during normal operation.
This placement keeps both the prepared execution environment and its references to shared GPU state outside the shared MPS failure domain.
When a fatal SM fault tears down the shared MPS context, the standby's context and memory mappings remain valid.
The standby can therefore take over using the retained state without repeating its initialization.

\MyPara{Post-Recovery Lifecycle Management.}
After a fatal SM fault, we restore every client through its standby, including the faulting client.
The recovered processes initially run outside MPS in independent CUDA contexts and time-share the GPU.
This temporary mode provides an isolation window for the operator to diagnose the faulting client.
Any repeated fault from that client remains confined to its own context and does not affect the other recovered clients.
We retain this mode until the operator terminates the faulting client.
We then launch replacement active processes for the surviving applications under MPS and transfer execution to them, restoring spatial sharing.
The surviving outside-MPS processes return to their standby roles, re-establishing the active--standby configuration for subsequent faults.
}

%% file: 06_Evaluation.tex
\section{Implementation and Evaluation}
\label{sec:evaluation}

\MyPara{Implementation.}
\revisionpass{
We implemented RoF and PoR with approximately 1000 lines of C in NVIDIA's open-source UVM module, covering RoF's fault-specific redirection paths and PoR's schedule-disable and pending-preemption sequence.
Our recovery substrate uses a roughly 500-line C build-time \texttt{libcuda.so.1} interceptor for VMM-based allocation sharing, together with user-space orchestration for the complete active--standby lifecycle.
We instantiate three application protocols for LLM serving, diffusion-based image generation, and ResNet50 image classification.
The LLM and diffusion protocols publish recovery state after each completed forward pass and denoising step\footnote{In the evaluated configurations, state publication reduces fault-free throughput by approximately 1\%.}, respectively, allowing non-faulting clients to resume from these application-defined boundaries. 
The ResNet50 protocol shares only model weights and requires no metadata synchronization.
Any memory corruption preceding the fault is confined to the faulting client's own state, which is discarded when the client is ultimately terminated by the operator.
To support characterization and evaluation, we also implement a deterministic fault-injection module covering all nine user-reachable MMU fault scenarios and five SM fault types.
Appendix~\ref{app:fault-injection} details the fault-injection methodology.
}

\revisionpass{
\MyPara{Evaluation Goals.}
Our evaluation mainly answers the following three questions:
(1) Does our isolation mechanism properly isolate all identified MMU fault types and preserve the process-level fail-stop semantics (\S\ref{sec:eval-isolation})?
\recovery{(2) Does our recovery substrate support recovery from SM faults across applications and GPU configurations effectively and efficiently (\S\ref{sec:eval-recovery})?}
(3) What overhead does our system incur during fault-free execution (\S\ref{sec:eval-overhead})?
}

\revisionpass{
\MyPara{Experimental setup.}
Unless otherwise noted, all experiments run on a single NVIDIA A6000 GPU (48\,GB VRAM) with an AMD EPYC 7763 CPU (60 cores) and 114\,GB DRAM, using driver version 580.95.05 and CUDA 13.0.
For end-to-end isolation and recovery, we run Qwen2.5-14B with vLLM.
We submit sequential batches of 128 ShareGPT requests (maximum 256 output tokens each), measuring aggregate output-token throughput at the server-side scheduler.
}

\revisionpass{
Our evaluation covers all nine user-reachable MMU fault scenarios and all five SM fault types.
We further test isolation with multi-stream execution, concurrent faulting threads, illegal-access loops, concurrent faulting clients, and multi-GPU configurations.
Recovery experiments span LLM serving, diffusion-based image generation, and ResNet50 image classification, including distributed LLM serving across multiple GPUs.
}

\subsection{Effectiveness of Fault Isolation}
\label{sec:eval-isolation}

\begin{table}[t]
\centering
\footnotesize
\setlength{\tabcolsep}{6pt}
\caption{MMU Fault containment results. 
All seven shared-TSG faults are successfully isolated, while the two per-client CE TSG faults remain confined by design.
}

\label{tab:isolation-correctness}
\begin{tabular}{c l c c c}
\toprule
\textbf{\#} & \textbf{Fault Type} & \textbf{Shared TSG?} & \textbf{No Isolation} & \textbf{Isolation} \\
\midrule
1  & OOB              & yes        & \textcolor{red}{\textbf{DIED}}  & \textcolor{green!50!black}{\textbf{ALIVE}} \\
2  & AM (CPU-res.)    & yes        & \textcolor{red}{\textbf{DIED}}  & \textcolor{green!50!black}{\textbf{ALIVE}} \\
3  & AM (GPU-res.)    & yes        & \textcolor{red}{\textbf{DIED}}  & \textcolor{green!50!black}{\textbf{ALIVE}} \\
4  & AM (VMM)         & yes        & \textcolor{red}{\textbf{DIED}}  & \textcolor{green!50!black}{\textbf{ALIVE}} \\
5  & Zombie           & yes        & \textcolor{red}{\textbf{DIED}}  & \textcolor{green!50!black}{\textbf{ALIVE}} \\
6  & Non-migr.        & yes        & \textcolor{red}{\textbf{DIED}}  & \textcolor{green!50!black}{\textbf{ALIVE}} \\
\midrule
7  & OOB (CE)         & per-client & \textcolor{green!50!black}{\textbf{ALIVE}} & \textcolor{green!50!black}{\textbf{ALIVE}} \\
8  & AM (CE)          & per-client & \textcolor{green!50!black}{\textbf{ALIVE}} & \textcolor{green!50!black}{\textbf{ALIVE}} \\
\midrule
11  & OOB (PBDMA)      & yes        & \textcolor{red}{\textbf{DIED}}  & \textcolor{green!50!black}{\textbf{ALIVE}} \\
\bottomrule
\end{tabular}

\end{table}

\MyPara{Fault Propagation Containment.}
We first validate that our isolation mechanism prevents MMU faults in one MPS client from propagating to a co-located client.
We co-locate two MPS clients: Client A runs our fault-injection module, while Client B repeatedly launches a kernel and checks for errors after each iteration.
We inject each of the nine user-reachable MMU fault combinations (\#1--\#8, \#11) into Client A and monitor Client B.
Table~\ref{tab:isolation-correctness} summarizes the results.
With isolation enabled, Client B survives all seven fault combinations that share the GR TSG.
The remaining two cases, CE OOB and CE AM, are already confined by per-client CE TSGs and leave Client B unaffected even without isolation, consistent with \S\ref{subsec:propagation}.
Overall, our mechanism isolates all reachable MMU faults.

\begin{figure}[t]
    \centering
    \includegraphics[width=\linewidth]{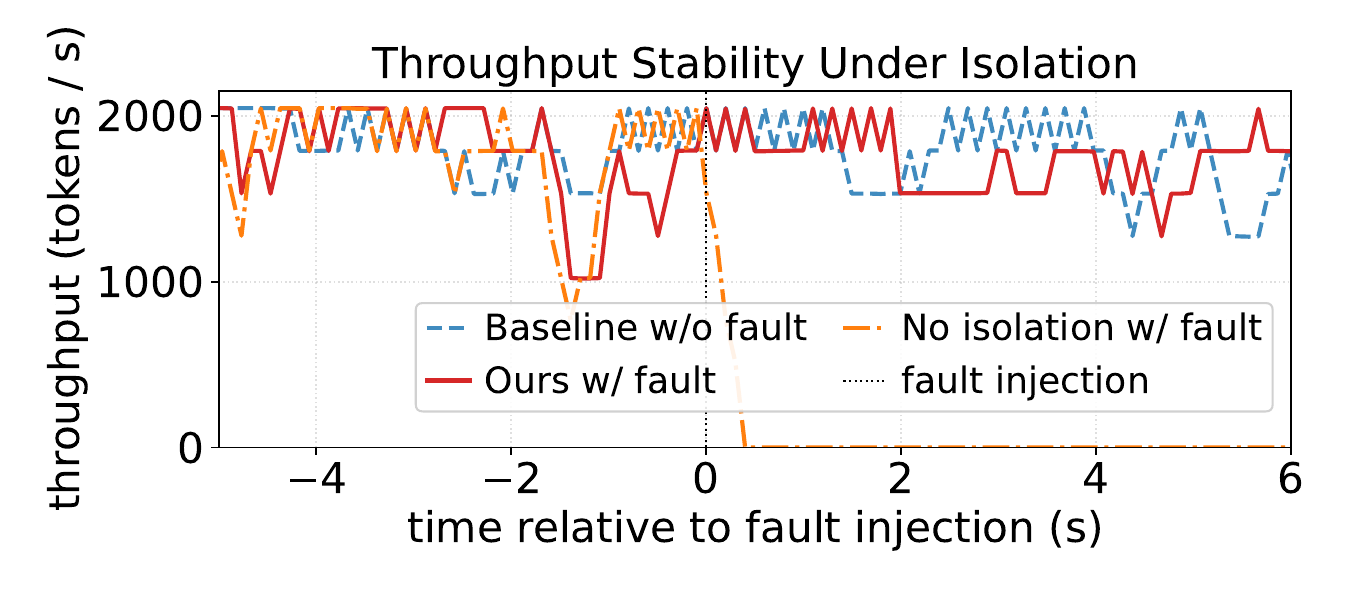}

    \caption{Throughput over time under fault isolation.}

    \label{fig:isolation-throughput}
\end{figure}

\MyPara{End-to-End Fault Isolation.}
\revisionpass{
We co-locate two MPS clients on the same GPU: Client A runs our fault-injection module, while Client B serves the LLM workload.
}
At $t=0$, we inject an SM OOB fault (\#1) into Client A.
Our mechanism intercepts the fault, redirects the invalid access to a dummy page, and terminates Client A, while Client B continues serving requests.
Figure~\ref{fig:isolation-throughput} reports Client B's throughput over time, along with no-fault and no-isolation baselines.
With isolation enabled, Client B's throughput matches the no-fault baseline, with no visible drop at the injection point, showing that fault handling is transparent to the co-running serving client.
In contrast, the no-isolation baseline crashes immediately after the injected fault.

\revisionpass{
\begin{table}[t]
    \centering
    \small
    \setlength{\tabcolsep}{4pt}
    \caption{Execution-witness outcomes.
    A value of 1 indicates an observed witness; 0 indicates none.}
    \label{tab:fail-stop}
    \begin{tabular}{@{}lccc@{}}
        \toprule
        Test case
        & \shortstack{Stock\\MPS}
        & \shortstack{RoF\\without PoR}
        & \shortstack{RoF\\with PoR} \\
        \midrule
        Single-stream, dependent   & 0 & 0 & 0 \\
        Single-stream, independent & 1 & 1 & 1 \\
        Multi-stream               & 0 & 1  & 0 \\
        \bottomrule
    \end{tabular}
\end{table}

\MyPara{Fail-Stop Semantics Preservation.}
Table~\ref{tab:fail-stop} compares stock MPS, RoF with MPS daemon termination, and RoF with PoR followed by the same termination procedure.
Across three repetitions of each single-stream SASS test, our full design matches stock MPS: the dependent witness remains zero, whereas the independent witness is one.
The independent witness does not require the faulting operation to complete, so its presence alone does not establish execution after fault resolution.
The multi-stream test monitors another independent stream in the same faulting process, which continues only under RoF without PoR.
PoR prevents this cross-stream continuation, preserving process-level fail-stop semantics during fault resolution.

\begin{table}[t]
    \centering
    \small
    \caption{Isolation across execution scenarios.
    Fault-to-safe measures time from fault occurrence to completed safe teardown of the faulting client.
    Ranges report the minimum and maximum latency across tested configurations.}
    \label{tab:isolation-robustness}
    \begin{tabular*}{\columnwidth}{@{\extracolsep{\fill}}lr@{}}
        \toprule
        Test configuration
        & \makecell[c]{Fault-to-safe(ms)} \\
        \midrule
        32/128/512 faulting threads
        & 113.8--182.7 \\
        \midrule
        \makecell[l]{Illegal-access loops}
        & 97.9--177.8 \\
        \midrule
        1/2/4/8 faulting clients
        & 186.7--1040 \\
        \midrule
        \makecell[l]{2-GPU LLM survivor;\\
                    1-GPU faulting client}
        & 263.1 \\
        \bottomrule
    \end{tabular*}
\end{table}

\MyPara{Isolation across Execution Scenarios.}
We evaluate four execution scenarios in Table~\ref{tab:isolation-robustness}.
Concurrent-thread kernels test whether isolation handles faults absent from the initial fault-buffer read; the 512-thread case processes 459 records across 38 rounds of fault servicing and replay before preemption completes.
Illegal-access loops test whether fault resolution permits further loop iterations; each run produces only one fault record and one replay, with dependent witnesses remaining zero.
Concurrent-client tests assess selective termination with multiple faulting processes: up to eight targets are terminated while a non-faulting client continues making progress.
The multi-GPU test checks whether isolating a single-GPU client preserves a co-located distributed workload; both workers of the two-GPU LLM remain healthy, and all 471 requests return correct results.
Across the tested configurations, all faulting clients are terminated and surviving clients remain functional.

}

\subsection{Effectiveness of Fault Recovery}
\label{sec:eval-recovery}

\begin{table}[t]
\centering
\footnotesize
\setlength{\tabcolsep}{8pt}
\caption{SM Fault recovery coverage. All five fault types trigger successful failover.}

\label{tab:recovery-coverage}
\begin{tabular}{c l c c}
\toprule
\textbf{\#} & \textbf{Fault Type} & \textbf{No Recovery} & \textbf{Recovery} \\
\midrule
15 & lane\_user\_stack\_overflow       & \textcolor{red}{\textbf{DIED}} & \textcolor{green!50!black}{\textbf{ALIVE}} \\
16 & illegal\_instruction       & \textcolor{red}{\textbf{DIED}} & \textcolor{green!50!black}{\textbf{ALIVE}} \\
17 & shared\_local\_oob & \textcolor{red}{\textbf{DIED}} & \textcolor{green!50!black}{\textbf{ALIVE}} \\
18 & misaligned\_addr                & \textcolor{red}{\textbf{DIED}} & \textcolor{green!50!black}{\textbf{ALIVE}} \\
19 & invalid\_addr\_space         & \textcolor{red}{\textbf{DIED}} & \textcolor{green!50!black}{\textbf{ALIVE}} \\
\bottomrule
\end{tabular}

\end{table}

\MyPara{Fault Recovery Coverage.}
We verify that our system can properly recover all SM faults. 
The experiment uses three processes: an active vLLM instance running as an MPS client, a standby instance idling outside MPS and ready to take over, and a co-located MPS client that triggers faults.
For each of the five SM faults, the fault-trigger process induces the fault, causing the shared MPS context to be destroyed and the active instance to terminate.
Because the standby runs outside MPS, it is isolated from this context destruction and remains available.
In all five cases, the standby detects the active failure and successfully takes over.
This result also demonstrates the benefit of our fault-agnostic recovery design: the standby relies on socket closure rather than fault-specific signals, so the same recovery path applies to any SM fault that terminates the active process.

\begin{figure}[t]
    \centering
    \includegraphics[width=\linewidth]{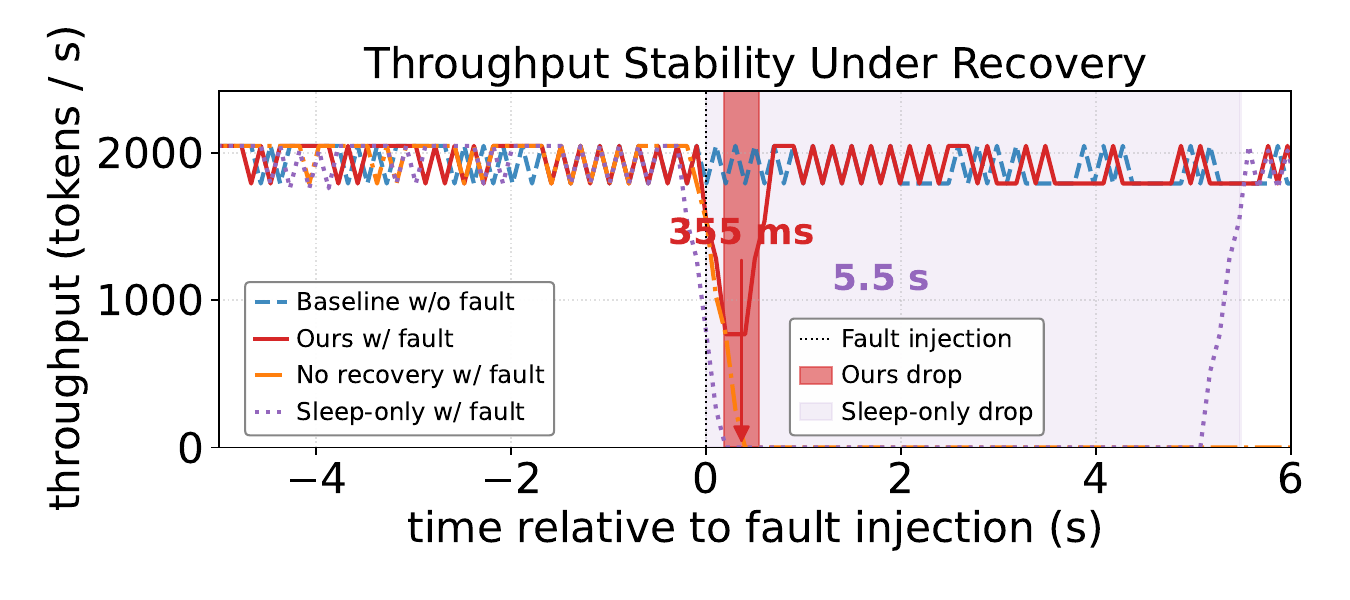}
    \caption{Throughput over time under fault recovery.}
    \label{fig:recovery-throughput}
\end{figure}

\MyPara{End-to-End Fault Recovery.}
\revisionpass{
We deploy an active--standby pair, with the active serving the LLM workload under MPS and the standby waiting outside MPS.
}
At $t=0$, we inject an illegal-instruction fault (\texttt{EXC\_4}) into the active instance, causing it to terminate and triggering failover.
Figure~\ref{fig:recovery-throughput} reports serving throughput over time.
We define the outage duration as the interval after the fault during which no tokens are produced.
With our mechanism, the outage lasts only 355\,ms, after which the standby resumes serving at the pre-fault throughput.
\revisionpass{
We construct a sleep-only baseline using vLLM's sleep mode.
It preserves the standby's initialized runtime but reloads model weights from CPU memory on failover, resulting in a 5.5\,s outage, over $15\times$ longer than our design.
}
Without recovery, throughput drops to zero permanently.
These results show that VMM-based recovery reduces service interruption from seconds to sub-second failover.

\revisionpass{
\begin{table}[t]
    \centering
    \small
    \setlength{\tabcolsep}{2pt}
    \caption{Fault-to-catchup latency and cold-restart breakdown.
    Fault-to-catchup measures time from fault occurrence until the application regains its pre-fault progress.}
    \label{tab:recovery-protocols}
    \begin{tabular*}{\columnwidth}
        {@{\extracolsep{\fill}}lrrrrr@{}}
        \toprule
        & & \multicolumn{4}{c}{Cold restart (s)} \\
        \cmidrule(l){3-6}
        Application
        & \makecell[c]{Recovery\\(ms)}
        & \makecell[c]{Init.}
        & \makecell[c]{State\\loading}
        & \makecell[c]{Progress\\reconstr.}
        & Total \\
        \midrule
        LLM (1 GPU)
        & 28.082 & 9.614 & 9.152 & 0.688 & 19.454 \\
        \midrule
        Diffusion
        & 628.521 & 4.052 & 8.684 & 4.627 & 17.363 \\
        \midrule
        ResNet50
        & 135.060 & 2.523 & 0.098 & 0.004 & 2.625 \\
        \midrule
        LLM (2 GPUs)
        & 48.35 & 15.662 & 4.605 & 0.768 & 21.035 \\
        \bottomrule
    \end{tabular*}
\end{table}
\MyPara{Recovery across Application Protocols.}
The single-GPU experiments in Table~\ref{tab:recovery-protocols} use an H100.
Both LLM configurations use Qwen2.5-14B.
The diffusion workload uses Qwen-Image to generate $1024\times1024$ images over 28 denoising steps.
We inject the fault to diffusion after the 14th denoising step completes.
State loading and progress reconstruction account for 51\% and 77\% of cold-restart latency for LLM serving and diffusion, respectively.
Our protocols reuse model weights to avoid reloading, and retain KV cache or denoising latents to avoid repeating completed computation.
For ResNet50, initialization accounts for 96\% of cold-restart latency, while weight loading and progress reconstruction contribute little.
Its protocol re-executes the interrupted request and primarily benefits from the prepared standby.
On two H100 GPUs, the LLM protocol resumes from its checkpoint and preserved KV cache with a fault-to-catchup latency of 48.35\,ms.
All 128 output tokens match the fault-free baseline.
}

\subsection{System Overhead}
\label{sec:eval-overhead}

\revisionpass{

\MyPara{Driver-Level Overhead.}
RoF and PoR execute only when handling fatal MMU faults and add no work to the normal execution path.
They therefore introduce no driver-level overhead during fault-free execution.

\MyPara{Standby Memory Overhead.}
VMM-based sharing avoids duplicating the GPU-resident state preserved for recovery.
The standby's additional GPU memory footprint comes from its initialized CUDA context and private runtime allocations.
For our implementation, this footprint is approximately 600\,MiB.
}

%% file: 07_Related_Works.tex
\section{Discussion: Static MPS SM Partitioning}
\label{sec:discussion}

\revisionpass{
Starting with driver r610, NVIDIA MPS provides partial isolation of SM-triggered faults when using static SM partitions~\cite{new-mps}.
Exclusive SM ownership enables the driver to associate SM error state with the owning partition.
This approach provides fault containment but restricts sharing flexibility: clients cannot borrow idle SMs from other partitions.
Our design does not require exclusive SM partitions: RoF and PoR isolate fatal MMU faults while allowing clients to share SMs, and the recovery substrate restores clients affected by fatal SM faults.
}

\section{Related Works}
\label{sec:related}


\revisionpass{
\MyPara{GPU Sharing Systems.}
GPU sharing systems improve utilization through workload scheduling~\cite{tgs,orion,gaiagpu} and resource partitioning.
Hardware partitioning with MIG offers physical isolation across GPU instances; Krypton~\cite{krypton} combines MIG partitioning with temporal sharing.
MPS-based systems such as LithOS~\cite{lithos}, GPUlet~\cite{gpulet}, GSLICE~\cite{gslice}, SGDRC~\cite{sgdrc}, and nvtaskset~\cite{nvtaskset} enable finer-grained spatial sharing through software-level resource control.
MuxFlow~\cite{muxflow} and LithOS~\cite{lithos} mitigate termination-induced fault propagation through graceful exit.
Such termination hazards can also be avoided using MPS's safe client-termination interface, even when the client has outstanding GPU work~\cite{mps-daemon}.
Our work addresses faults arising during GPU execution, providing MMU fault containment with process-level fail-stop semantics and fast recovery from SM faults.
}

\revisionpass{
\MyPara{GPU Fault Tolerance.}
Checkpointing systems preserve GPU application state~\cite{checkfreq,gemini,crac}, while distributed training systems exploit redundancy and reconfiguration~\cite{bamboo,oobleck,unicron}.
LLM serving systems provide workload-specific recovery mechanisms~\cite{tarragon,kevlarflow,dejavu,spotserve,serverlessllm}.
We target fatal SM faults that terminate MPS clients but leave the GPU operational.
Reusing application-designated GPU state and pre-initialized standbys avoids state reloading, reconstruction of preserved progress, and initialization on the recovery critical path.
Our substrate supports application-specific protocols and a complete, repeatable recovery lifecycle.
}

%% file: 08_Discussion_and_Conclusion.tex
\section{Conclusion}

This work designs fault-resilient MPS.
We first presented a systematic characterization of GPU faults under MPS, classifying 19 fault scenarios into three categories and identifying the architectural boundaries between isolable and non-isolable faults.
\revisionpass{
Guided by this characterization, we designed two complementary mechanisms: driver-level isolation that contains all reachable MMU faults while preserving process-level fail-stop semantics, and VMM-based active--standby recovery for SM faults.
In an end-to-end case study, isolation incurs no visible outage, while recovery restores pre-fault throughput in 355\,ms.
}
Together, these mechanisms bring fault resilience to fine-grained GPU sharing under MPS.

%% file: appendix.tex
\section{Fault Characterization and Injection}
\label{app:implementation}

\subsection{Fault Injection Module}
\label{app:fault-injection}

Our fault-injection module provides deterministic triggers for the nine reachable MMU fault scenarios and five fatal SM fault types used in our characterization and evaluation.
Table~\ref{tab:fault-injection} summarizes their trigger methods and faulting engines.

\MyPara{MMU Fault Injection.}
Seven MMU fault scenarios are triggered through CUDA API programs.
For zombie-range and non-migratable-range accesses, we use UVM debug ioctls to establish the corresponding allocation states before a kernel accesses them.
Table~\ref{tab:memory-fault-taxonomy} distinguishes these two debug-ioctl cases from those triggered through CUDA APIs.

\MyPara{SM Fault Identification and Injection.}
We systematically examine the \texttt{CUDA\_EXCEPTION} codes documented in the CUDA-GDB manual~\cite{cuda_gdb}, construct trigger programs, and classify the resulting faults by their driver-handling paths.
Faults that invoke UVM fault processing belong to the MMU category.
In contrast, SM faults trigger RC recovery through RM/GSP without entering UVM fault processing.
Some legacy exception codes remain documented but cannot be triggered on the evaluated GPUs.
The \texttt{warp\_assert} case also does not trigger RC recovery in our tests and is excluded from the fatal SM fault set.
This procedure identifies the five distinct fatal SM fault types reported in Table~\ref{tab:memory-fault-taxonomy}.
We further cross-check the exception classification against Nouveau, the Linux community's reverse-engineered NVIDIA GPU driver.

\begin{table*}[t]
    \centering
    \small
    \setlength{\tabcolsep}{4pt}
    \caption{Fault-injection methods, with fault numbers matching Table~\ref{tab:memory-fault-taxonomy}.
    OOB denotes out-of-bounds access; AM denotes access mismatch.}
    \label{tab:fault-injection}
    \begin{tabular*}{\textwidth}
        {@{\extracolsep{\fill}}r p{0.22\textwidth} p{0.60\textwidth} l@{}}
        \toprule
        \textbf{\#}
        & \textbf{Fault Scenario}
        & \textbf{Trigger Method}
        & \makecell[c]{\textbf{Faulting}\\\textbf{Engine}} \\
        \midrule
        \multicolumn{4}{@{}l}{\textbf{MMU Faults}} \\
        \midrule
        1
        & OOB
        & \code{cudaMalloc} followed by a kernel write to an unmapped address beyond the allocation
        & SM \\
        \cmidrule(lr){1-4}
        2
        & AM (CPU-resident)
        & Kernel write to a CPU-resident range with incompatible access permissions
        & SM \\
        \cmidrule(lr){1-4}
        3
        & AM (GPU-resident)
        & Kernel write to a GPU-resident range with incompatible access permissions
        & SM \\
        \cmidrule(lr){1-4}
        4
        & AM (VMM)
        & VMM allocation and mapping with read-only access set through \code{cuMemSetAccess}, followed by a kernel write
        & SM \\
        \cmidrule(lr){1-4}
        5
        & Zombie range
        & UVM debug ioctl to invalidate the range's backing, followed by a kernel access
        & SM \\
        \cmidrule(lr){1-4}
        6
        & Non-migratable range
        & UVM debug ioctl to prohibit migration from host memory, followed by a kernel access
        & SM \\
        \cmidrule(lr){1-4}
        7
        & OOB
        & \code{cuMemcpy} targeting an unmapped address beyond an allocation
        & CE \\
        \cmidrule(lr){1-4}
        8
        & AM
        & Copy-engine write to a destination with incompatible access permissions
        & CE \\
        \cmidrule(lr){1-4}
        11
        & OOB
        & \code{cuStreamWaitValue32} on an unmapped virtual address
        & PBDMA \\
        \midrule
        \multicolumn{4}{@{}l}{\textbf{SM Faults}} \\
        \midrule
        15
        & Lane user stack overflow
        & Deep recursion with the CUDA stack-size limit set to 1\,KB
        & SM \\
        \cmidrule(lr){1-4}
        16
        & Illegal instruction
        & Driver-API launch of a patched cubin containing an invalid opcode
        & SM \\
        \cmidrule(lr){1-4}
        17
        & Shared/local OOB
        & Inline PTX \code{ld.shared} or \code{ld.local} targeting an out-of-bounds address
        & SM \\
        \cmidrule(lr){1-4}
        18
        & Misaligned address
        & Misaligned global-memory access
        & SM \\
        \cmidrule(lr){1-4}
        19
        & Invalid address space
        & \code{atom.global.add} targeting a shared-memory address
        & SM \\
        \bottomrule
    \end{tabular*}
\end{table*}

\subsection{MMU Fault Catalog}
\label{sec:appendix-faults}

This subsection details the MMU fault conditions and reachability boundaries summarized in \S\ref{subsec:taxonomy}.
Fault numbers refer to Table~\ref{tab:memory-fault-taxonomy}.

\subsubsection{Parse-Time Fatal Faults}
\label{sec:appendix-parsetime}
Parse-time fatal faults are identified during initial fault parsing, before UVM attempts any resolution.
These correspond to errors that cannot be resolved through any software intervention.
UVM defines 12 such fault types across five categories:
(1)~\emph{MMU structural corruption} (e.g., malformed page directory entries, virtual address limit violations);
(2)~\emph{Context/channel state corruption} (e.g., unbound instance blocks, corrupted command streams);
(3)~\emph{Privilege/security violations} (e.g., unprivileged SM instructions, confidential computing boundary violations);
(4)~\emph{Memory attribute/aperture mismatch} (e.g., unsupported apertures);
(5)~\emph{Hardware data corruption} (ECC-poisoned pages).

\subsubsection{Deferred-to-Servicing Fault Scenarios}
\label{sec:appendix-deferred}

Faults that pass the parse-time check enter UVM's servicing pipeline, which examines the target VA range, access permissions, and available migration or mapping operations.
The non-serviceable cases fall into four conditions.

\MyPara{Out-of-Bounds Access.}
The faulting virtual address falls outside the VA ranges registered with UVM (\#1, \#7, and \#11).
Because no registered range owns the address, UVM has no backing allocation with which to service the access.

\MyPara{Access Mismatch.}
The requested access conflicts with the target range's permissions.
For SM accesses, we distinguish CPU-resident (\#2), GPU-resident (\#3), and VMM external-range (\#4) cases according to page residency or range type.
In the CPU- and GPU-resident cases, migrating the page alone cannot repair the incompatible permissions.
The VMM case instead involves an externally managed range whose mappings cannot be repaired through UVM's managed-memory servicing path.
The taxonomy also includes access mismatch originating from the copy engine (\#8).

\MyPara{Zombie Range Access.}
The access targets a range whose backing has been invalidated during teardown, while the range remains present in UVM's bookkeeping (\#5).
UVM therefore finds the range but cannot service the access using valid backing.

\MyPara{Non-Migratable Range Access.}
The target allocation is bound to a processor from which migration is prohibited (\#6).
If servicing the access requires that migration, UVM cannot resolve the fault.

\subsubsection{Unreachable Combinations}
\label{sec:appendix-unreachable}

The remaining five engine--fault combinations are not exposed by the CUDA API paths evaluated in our characterization.

\MyPara{CE Zombie Range (\#9) and CE Non-Migratable Range (\#10).}
Zombie range and non-migratable range faults arise exclusively from the lifecycle of UVM-tracked allocations (deallocation timing and migration policy, respectively).
The CUDA runtime dispatches all user operations on managed memory as SM kernels.
Consequently, the Copy Engine never encounters these allocation states through normal CUDA API usage.

\MyPara{PBDMA Access Mismatch (\#12), Zombie Range (\#13), and Non-Migratable Range (\#14).}
The PBDMA semaphore API rejects managed memory at the API layer, preventing PBDMA faults from targeting user-provided virtual addresses.
PBDMA faults instead target an internal semaphore pool address managed by the driver.
As a result, the only PBDMA fault reachable from user space is out-of-bounds access (\#11), which can occur when a semaphore operation references an invalid pool offset.